\documentclass[twocolumn,showpacs,floatfix,preprintnumbers,amsmath,amssymb,prl]{revtex4}

\usepackage[dvips]{graphicx}    % Include figure files
\usepackage{epsfig}      % Include figure files
\usepackage{dcolumn}     % Align table columns on decimal point
\usepackage{bm}          % Bold math

\def\lamno{LaMnO$_3$}
\def\mnoVI{MnO$_6$}

\def\oprime{$O^\prime$}
\def\mn#1{Mn$^{#1}$}
\newcommand{\etal}{{\it et al.}}

\def\im3{{Im\overline 3}}
\def\tjt{$T_{JT}$}
\def\rmnoave{$\langle r_{Mn-O}\rangle$}

\newcommand{\degree}{$^\circ$C }
\newcommand{\bigbullet}{\mbox{\LARGE $\bullet$}}
\newcommand{\rmax}{$r_{max}$}

\newcommand{\rwp}{R$_{wp}$}

\def\sb#1{$_{#1}$}

\begin{document}

\title{Orbital correlations in the pseudo-cubic \emph{O}
and rhombohedral {\boldmath ${R}$}-phases of \lamno}

\author{Xiangyun Qiu,$^1$ Th.~Proffen,$^2$ J.~F.~Mitchell,$^3$ and S.~J.~L.~Billinge$^1$}
\affiliation{$^1$Department of Physics and Astronomy,  Michigan
State University, E.~Lansing, MI 48824\\ $^2$Los Alamos National
Laboratory, LANSCE-12, MS H805, Los Alamos, NM,
87545\\ $^3$Material Science Division, Argonne National Laboratory,
Argonne, IL, 60439}

\date{\today}

\begin{abstract}
The local and intermediate structure of stoichiometric \lamno\ has
been studied in the pseudocubic and rhombohedral phases at high
temperatures (300 to 1150~K). Neutron powder diffraction data were
collected and a combined Rietveld and high real space resolution
atomic pair distribution function analysis carried out. The nature
of the Jahn-Teller (JT) transition around 750~K is confirmed to be
orbital order to disorder. In the high temperature orthorhombic
($O$) and rhombohedral ($R$) phases the  \mnoVI\ octahedra are still
fully distorted locally. The data suggest the presence of local
orbitally ordered clusters of diameter $\sim 16$~\AA\ ($\sim$four
\mnoVI\ octahedra) implying strong nearest neighbor JT
anti-ferrodistortive coupling.
\end{abstract}
%61.12.-q   Neutron diffraction and scattering
%75.47.Gk   Colossal magnetoresistance
%75.47.Lx   Manganites
\maketitle
% general and short introduction of manganites
The perovskite manganites related to \lamno\ continue to yield
puzzling and surprising results despite intensive study since the
1950's~\cite{ramir;jpcm97,salam;rmp01,dagot;b;npsacm}.  The
pervading interest comes from the delicate balance between
electronic, spin and lattice degrees of freedom coupled with strong
electron correlations. Remarkably, controversy still exists about
the nature of the undoped endmember material, \lamno , where every
manganese ion has a nominal charge of 3+ and no hole doping exists.
The ground-state is well understood as an A-type antiferromagnet
with long-range ordered, Jahn-Teller (JT) distorted, MnO\sb{6}
octahedra~\cite{wolla;pr55,rodri;prb98} that have four shorter and
two longer bonds~\cite{proff;prb99}.  The elongated occupied $e_g$
orbitals lie down in the $xy$-plane and alternate between pointing
along $x$ and $y$ directions, the so-called \oprime\ structural
phase~\cite{wolla;pr55,rodri;prb98}.  At $T_{JT}\sim 750$~K the
sample has a first-order structural phase transition to the $O$
phase that formally retains the same symmetry but is pseudo-cubic
with almost regular MnO\sb{6} octahedra (six almost equal
bond-lengths).  In this phase the cooperative JT distortion has
essentially disappeared.  It is the nature of this $O$ phase that is
unclear.  The $O$-phase has special importance since it is the phase
from which ferromagnetism and colossal magnetoresistance appears at
low temperature at Ca, Sr dopings $>
0.2$~\cite{wolla;pr55,ramir;jpcm97}.

The additional complication of disorder due to the presence of
alkali-earth dopant ions, and the doped-holes on manganese sites, is
absent in \lamno . However, there is still disagreement about the
precise nature of the $O$-phase in this simple case. In 1996
Millis~\cite{milli;prb96}, based on fits to data of a classical
model that included JT and lattice terms, estimated the energy of
the JT splitting to be $\gtrsim 0.4$~eV and possibly as high as
$2.4$~eV. This high energy scale clearly implies that the JT
distortions are not destroyed by thermal excitation at 750~K and the
\oprime - $O$ transition is an order-disorder transition where local
JT octahedra survive but lose their long-range spatial correlations.
%The barrier that is overcome at this transition is the lattice
%%stiffness that serves to orientationally order the distorted
%octahedra.
This picture is supported by some probes sensitive to
local structure, for example XAFS~\cite{araya;jmmm01,sanch;prl03}
and Raman scattering~\cite{grana;prb00}, each of which presents
evidence that structural distortions consistent with local JT
effects survive above $T_{JT}$.  However while compelling, this
picture seems hard to reconcile with the electronic and magnetic
properties.  In the $O$-phase the material's conductivity increases
(despite being at higher temperature and more disordered), and
becomes rather temperature
independent~\cite{manda;prb01,zhou;prb99}, and the ferromagnetic
correlations become stronger compared to the
\oprime-phase~\cite{zhou;prb99,tovar;prb99}. These results clearly
imply greater electronic mobility in the $O$-phase, which appears at
odds with the persistence of JT distortions above $T_{JT}$ that are
becoming orientationally disordered.  The disordering should result
in more carrier scattering and higher resistivity, contrary to the
observation~\cite{zhou;prb99}.  Furthermore, a sharp decrease in the
unit cell volume at $T_{JT}$ has been reported~\cite{chatt;prb03}.
This mimics the behavior observed when electrons delocalize and
become mobile at $T_c$ in the doped materials~\cite{radae;prb97},
whereas from geometrical arguments, an orbital disordering of rigid
octahedra would result in a volume increase.
%The authors of Ref.~\cite{chatt;prb03} reconcile this observation with
%the order-disorder picture of the phase transition by analogy with the
%volume drop on melting of ice, though the connection seems somewhat
%tenuous.

We have used a diffraction probe of the local atomic structure,
neutron pair distribution function (PDF) analysis, to see if the local
and average behaviors can be reconciled and understood.  This method
allows quantitative structural refinements to be carried out on
intermediate length-scales in the nanometer range.  We confirm that
the JT distortions persist locally at all temperatures, and quantify
the amplitude of these \emph{local} JT distortions. No local
structural study exists of the high-temperature rhombohedral R-phase
that exists above $T_R=1010$~K. We show for the first time that even
in this phase where the octahedra are constrained by symmetry to be
undistorted, the local JT splitting persists. Fits of the PDF over
different $r$-ranges indicate that locally distorted domains have a
diameter of $\sim 16$~\AA\ with about four MnO\sb{6} octahedra
spanning the locally ordered cluster.
%The unit cell
%volume collapse seen crystallographically comes from a reduction in
%the average Mn-O bond-length, \ud{which, however,} is constant with
%temperature in the local structure.
We show that the crystallographically observed volume reduction at
\tjt~\cite{chatt;prb03} is consistent with increased octahedral
rotational degrees of freedom and is not the reduction in \mnoVI\
observed crystallographically. We speculate that nanoclusters of the
low-T structure form on cooling at $T_R$ but are disordered and align
themselves along the three crystal axes. These domains grow, and the
orbital ordering pattern becomes long-range ordered below
\tjt .
%\section{EXPERIMENTS}

Powder samples of $\sim 6$~g were prepared from high purity MnO$_2$
and La$_2$O$_3$; the latter was pre-fired at 1000 \degree\ to remove
moisture and carbon dioxide. Final firing conditions were chosen to
optimize the oxygen stoichiometry at 3.00. The crystallographic
behavior was confirmed by Rietveld refinements
%of the same data used
%in the PDF,
using program GSAS~\cite{larso;unpub87}.  Both the differential
thermal analysis (DTA) and Rietveld measurements estimated the same
phase transition temperatures of \tjt $\sim$ 735~K and $T_R \sim$
1010~K. These values indicate that the sample is highly
stoichiometric~\cite{norby;jssc95,rodri;prb98}.
%, e.g.\ \tjt=750~K for perfectly stoichiometric sample
%and around \tjt$\sim$ 600~K reported for 0.005 excess oxygen per
%chemical formula unit
Neutron powder diffraction data were collected on the NPDF
diffractometer at the Lujan Center at Los Alamos National
Laboratory. The sample, sealed in a cylindrical vanadium tube, was
measured from 300 to 1150~K.  After temperature cycling, a further
measurement at 300~K confirmed that the sample was unchanged by the
thermal cycling in reducing atmosphere.
%The sample stoichiometry was then
%verified to remain the same by a later measurement at 300~K.}
% The data
% were corrected for detector deadtime and efficiency, background,
% absorption, multiple scattering, inelasticity effects and normalized
% by the incident flux and the total sample scattering cross-section to
% yield the total scattering structure function, $S(Q)$, where $Q$ is
% the magnitude of the scattering vector. The PDF,
% \gr\ is then obtained by a Fourier transform according to \sqtogr.
% %
%\begin{equation}
%  G(r)= \frac{2}{\pi }\int_{0}^{\infty} Q [S(Q)-1] \sin (Qr)\> dQ.
%\end{equation}
%
Data reduction to obtain the PDFs~\cite{egami;b;utbp03} was carried
out using program PDFgetN~\cite{peter;jac00}. PDF Modeling was carried
out using the program PDFFIT~\cite{proff;jac99}.

%Data were also collected at the SEPD diffractometer at the Intense
%Pulsed Neutron Source (IPNS) at Argonne National Laboratory on a
%different sample that was the subject of an earlier low-temperature
%study~\cite{proff;prb99}.  This sample had a slight oxygen
%non-stoichiometry, as evidenced by lowered structural transition
%temperatures.  The PDF results from this sample were qualitatively
%\emph{the same} as from the stoichiometric sample presented in this
%letter.
%
%Experimental data and the model fitting at 20~K are shown in
%Fig.~\ref{fig;soq}.

\begin{figure}[tbp]
  \centering \includegraphics[width=0.46\textwidth]{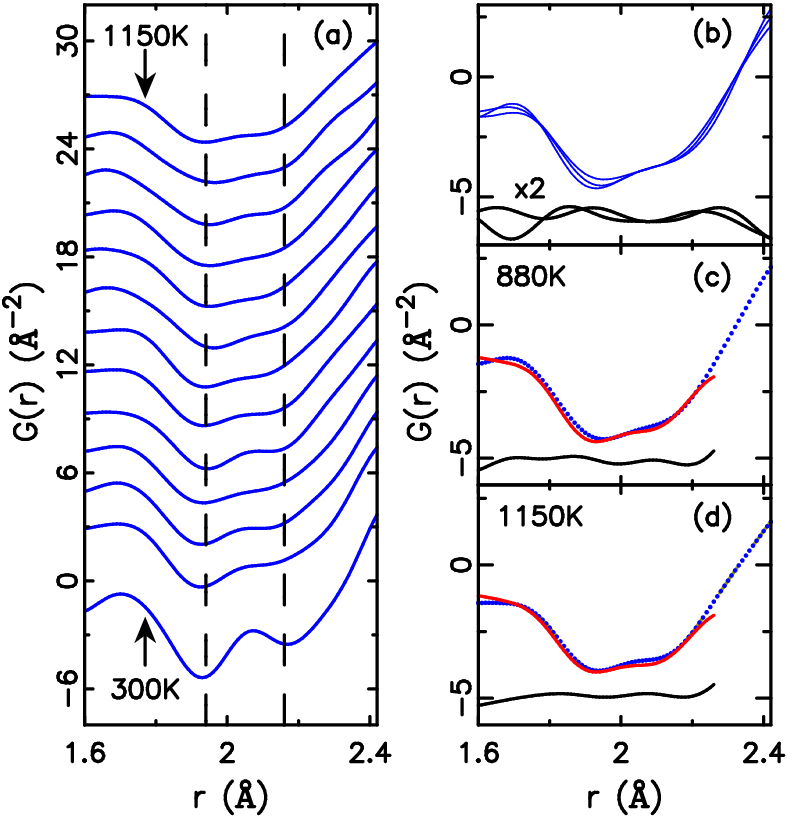}
  \caption{(a) Low-$r$ region of the experimental PDFs at all
  temperatures offset along the $y$ axis with increasing
  temperature. Two dashed lines indicate the Mn-O short and long bonds
  at 1.94 and 2.16 \AA, respectively. (b) PDFs at 650, 720, and 880~K
  without offset. The differences between 720 and both 650~K and 880~K
  (both high T minus low T) are shown offset below. (c) Two-Gaussian
  fit of PDF data at 880~K (solid dots).  Solid line denotes fitted
  curve with the difference curve offset below.  (d) Same as (c) for
  1150~K PDF data.}  \label{fig;lowrpdf}
\end{figure}

%
We first confirm the earlier XAFS
results~\cite{araya;jmmm01,sanch;prl03} and establish that in the
$O$-phase the full JT distortion persists in the local structure.
This result can be seen qualitatively in Fig.~\ref{fig;lowrpdf}.
%%%%%%%%%%%%%%%%%%%%%%%%%%%%%%%%%%%%%%%%%%%%%%%%%%%%%%%%%%%%%%%%%%%%
%
%  Here FIGURE starts ...
%%%%%%%%%%%%%%%%%%%%%%%%%%%%%%%%%%%%%%%%%%%%%%%%%%%%%%%%%%%%%%%%%%%
%
%
%%%%%%%%%%%%%%%%%%%%%%%%%%%%%%%%%%%%%%%%%%%%%%%%%%%%%%%%%%%%%%%%%%%
%
%  Here FIGURE ends!
%%%%%%%%%%%%%%%%%%%%%%%%%%%%%%%%%%%%%%%%%%%%%%%%%%%%%%%%%%%%%%%%%%%
%
In panel (a) the first peak in the PDF, coming from Mn-O nearest
neighbor (nn) bonds, is shown.  This peak is upside-down because of
the negative neutron scattering length of Mn~\cite{egami;b;utbp03}.
It is double-valued because of the presence of short-bonds of average
length 1.94~\AA, and long-bonds of length 2.16~\AA , coming from the
JT distorted octahedra.  The doublet is clearly evident at low
temperature below \tjt.  However, the splitting persists up to the
highest temperature and remains larger than the thermal broadening of
the peaks.  This shows qualitatively and intuitively that the JT
distortion survives at all temperatures. The peaks do broaden at
higher temperature due to increased thermal
motion~\cite{egami;b;utbp03} and the two contributions to the doublet
are not resolved at high temperature. However, it is clear that the
peak is not a broad, single-valued, Gaussian at high temperature as
predicted from the crystallographic models.

We would like to see if there is any change in the peak profile on
crossing the \tjt\ beyond normal thermal broadening.  To test this
we plot the change in this nn Mn-O peak between T=720~K and T=880~K,
as it crosses \tjt, and compare this to the change in the peak on
going from 650~K to 720~K. The latter case has approximately the
same temperature differential (thermal broadening will be
comparable) but there is no structural transition in that range. The
difference curves from these two situations, shown in
Fig.~\ref{fig;lowrpdf}(b), are almost identical.  We conclude that
there is virtually no change to the MnO\sb{6} octahedra as they go
from the \oprime\ to the $O$ phase.

\begin{figure}[tbp]
  \centering \includegraphics[width=0.46\textwidth]{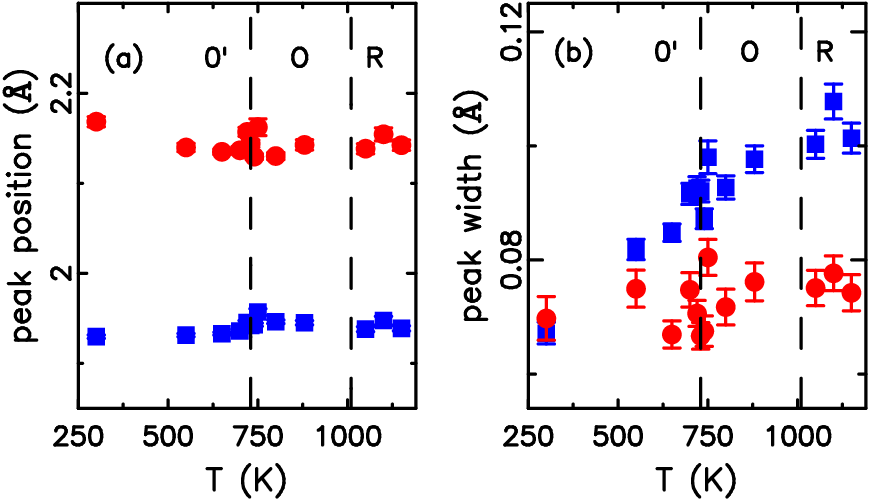}
  \caption{Fitting results of the Mn-O PDF peaks with two-Gaussians:
  the low-$r$ peak ($\blacksquare$) around 1.94 \AA, and the high-$r$
  peak (\bigbullet) $\sim$ 2.16 \AA. The dashed lines denote the phase
  transition temperatures. (a) Peak positions. (b) Peak widths. } 
  \label{fig;peakpar}
\end{figure}

Finally, to quantify the nature of the local JT distortions, we have
fit two Gaussian peaks to the nearest-neighbor \mnoVI\ doublet in the
experimental data at all temperatures.  The quality of the fits at
``low'' (880~K) and high (1150~K) temperature are rather good, as can
be seen in Fig.~\ref{fig;lowrpdf}(c) and (d). The position and width
of each peak, and the total integrated intensity of the doublet, were
allowed to vary (the intensities of the long and short bond
distributions were constrained to be 1:2 in the fits). The weighted
residual factor \rwp\ varies from 1\% at high temperature to 5\% at
low-temperature indicating excellent agreement over all
temperatures. Attempts to fit a single Gaussian in the $O$ phase
resulted in significantly worse agreements.  The results of the
temperature-dependent fits are shown in Fig.~\ref{fig;peakpar}.
%%%%%%%%%%%%%%%%%%%%%%%%%%%%%%%%%%%%%%%%%%%%%%%%%%%%%%%%%%%%%%%%%%%%
%
%  Here FIGURE starts ...
%%%%%%%%%%%%%%%%%%%%%%%%%%%%%%%%%%%%%%%%%%%%%%%%%%%%%%%%%%%%%%%%%%%
%
%
%%%%%%%%%%%%%%%%%%%%%%%%%%%%%%%%%%%%%%%%%%%%%%%%%%%%%%%%%%%%%%%%%%%
%
%  Here FIGURE ends!
%%%%%%%%%%%%%%%%%%%%%%%%%%%%%%%%%%%%%%%%%%%%%%%%%%%%%%%%%%%%%%%%%%%
No anomaly of any kind can be identified across the \tjt=735~K. The
integrated intensities and peak positions do not change
significantly with temperature from 300~K to 1150~K indicating that
the average bond-lengths of the short- and long-bonds in the JT
distorted octahedra are rather temperature independent.  The peak
widths increase slightly with temperature due to increased thermal
motion (Fig.~\ref{fig;peakpar}(b)); however, there is no clear
discontinuity in the peak broadening associated with the phase
transition. The results confirm that the {\it local} JT distortions
persist, virtually unstrained, into the $O$ and $R$-phases.

\begin{figure}[tbp]
  \centering \includegraphics[width=0.46\textwidth]{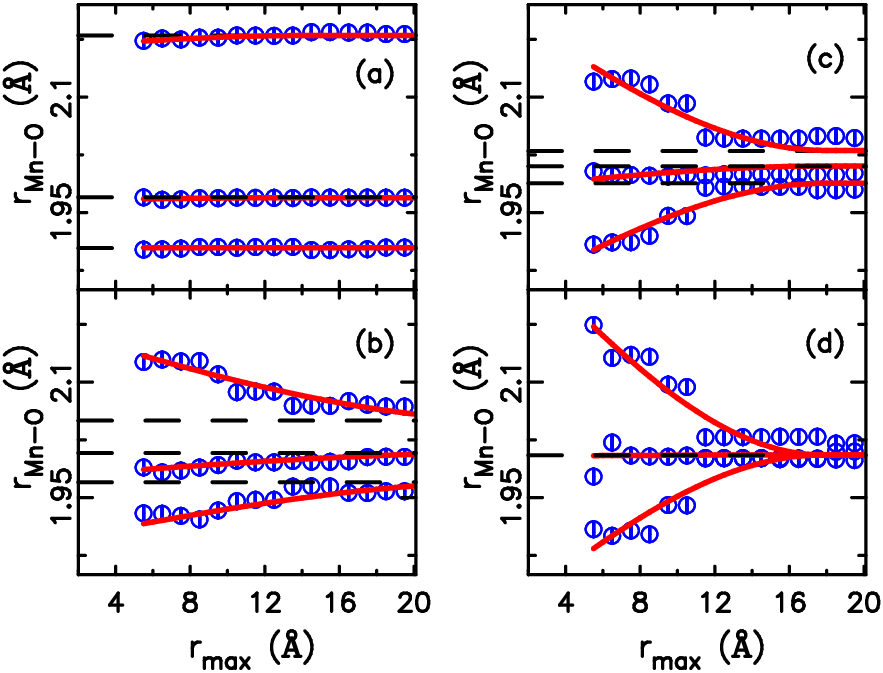}
  \caption{Mn-O bond lengths from the refined structure model as a
  function of $r_{max}$ are shown as circles. Error bars are smaller
  than, but comparable, to the symbol size.  The solid red lines are
  the expected behavior assuming spherical locally ordered
  clusters. All three bond lengths are self-consistently fit with a
  single parameter, the cluster size.  The horizontal dashed lines
  indicate the Mn-O bond lengths from Rietveld refinements. (a) 300~K,
  (b) 740~K, (c) 800~K, (d) 1100~K.} \label{fig;varyrangefit}
\end{figure}

\begin{figure}[tbp]
  \centering \includegraphics[width=0.46\textwidth]{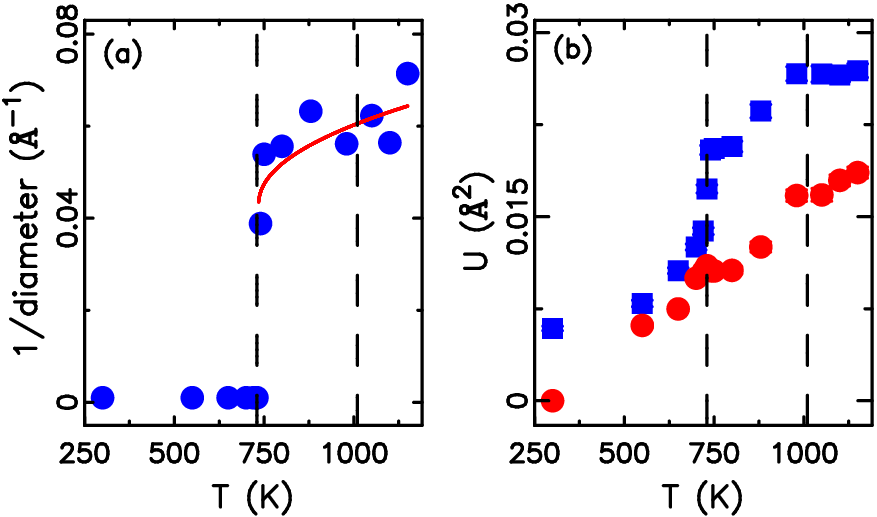}
  \caption{ (a) Inverse of the domain diameter obtained from the fits
  shown in Fig.~\ref{fig;varyrangefit}. Solid is a simple exponential
  critical curve fit as a guide to the eye.  (b) Thermal displacement
  parameters of oxygen atoms parallel (blue $\blacksquare$) and
  perpendicular (red \bigbullet) to Mn-O bonding directions. The
  perpendicular component is shift down by 0.007 \AA$^2$ for
  clarity. } \label{fig;ouaniso}
\end{figure}

Fits to the nearest neighbor Mn-O peaks show a large JT distortion in
the $O$-phase, whereas crystallographically the \mnoVI\ octahedra are
almost regular (six equal bonds).  This implies that the average
structure results from a loss of coherence of the ordering of the JT
distorted octahedra and that the JT transition is an orbital
order-disorder transition.  In principle we can test the extent of any
orbital short-range-order in the PDF by refining models to the data
over different ranges of~$r$.  Fits confined to the low-$r$ region
will yield the local (JT distorted) structure whereas fits over wider
ranges of $r$ will gradually cross over to the average
crystallographic structure.  To extract the size of the short-range
ordered clusters, we have fit the PDF from $r_{min}$=$1.5$ \AA\ to
\rmax , where $r_{max}$ was increased step by step from 5~\AA\ to
20~\AA , by which time the PDF refinement agrees with the average
structure refinement from Rietveld. This was done for all
data-sets. The model used at all temperatures was the low-temperature
structure in the $Pbnm$ space group. Representative results are shown
in Fig.~\ref{fig;varyrangefit}.  Note that three distinct bond-lengths
are obtained from modeling although only two peaks can be resolved
directly in the PDF at low-$r$.
%%%%%%%%%%%%%%%%%%%%%%%%%%%%%%%%%%%%%%%%%%%%%%%%%%%%%%%%%%%%%%%%%%%%
%
%  Here FIGURE starts ...
%%%%%%%%%%%%%%%%%%%%%%%%%%%%%%%%%%%%%%%%%%%%%%%%%%%%%%%%%%%%%%%%%%%
%
%
%%%%%%%%%%%%%%%%%%%%%%%%%%%%%%%%%%%%%%%%%%%%%%%%%%%%%%%%%%%%%%%%%%%
%
%  Here FIGURE ends!
%%%%%%%%%%%%%%%%%%%%%%%%%%%%%%%%%%%%%%%%%%%%%%%%%%%%%%%%%%%%%%%%%%%
The amplitude of the refined JT distortion is constant as a function
of \rmax\ at 300~K reflecting the fact the orbital order is
perfectly long-range. Regardless of the range fit over, the full JT
distortion is recovered (Fig.~\ref{fig;varyrangefit}(a)).  At higher
temperature, and especially in the $O$ phase the amplitude of the
refined distortion falls off smoothly as the fit range is extended
to higher-$r$, until it asymptotically approaches the much smaller
crystallographically refined value.  We understand this behavior in
the following way.  Domains of local orbital order exist in the $O$
phase.  These may resemble the pattern of orbital order at
low-temperature (\oprime\ phase), and for convenience this is how we
have modeled them.  These domains do not propagate over long-range
and are orientationally disordered in such a way that, on average,
the observed pseudo-cubic structure is recovered. We can estimate
the domain size by assuming that the orbitals are ordered inside the
domain but uncorrelated from one domain to the neighboring domain.
This results in a fall-off in the amplitude of the refined
distortion with increasing fit range with a well defined PDF
form-factor~\cite{qiu;unpub04}, assuming spherical domains, that
depends only on the diameter of the domain.  The three curves of
refined-bond-length vs.\ \rmax\ from the short, medium and long
bonds of the \mnoVI\ octahedron could be fit at each temperature
with the diameter of the domain as the \emph{one single} parameter.
Representative fits are shown in Fig.~\ref{fig;varyrangefit} as the
filled circles.  The temperature dependence of the inverse domain
diameter is shown in Fig.~\ref{fig;ouaniso}(a). We find that in the
pseudo-cubic $O$-phase these clusters have a diameter of $\sim
16$~\AA , roughly independent of temperature except close to \tjt\
where the size grows. Below \tjt\ the correlation length of the
order is much greater, although some precursor effects are evident
just below \tjt. The refined domain size of orbital order is similar
in the \emph{R}-phase though the quality of the fits becomes worse
in this region.  This may be because the nature of the short-range
orbital correlations changes, i.e., becomes different from the
\oprime\ phase. The correlation length scale of 16 \AA\ spans over
four \mnoVI\ octahedra, suggesting strong nearest neighbor JT
antiferrodistortive coupling (as expected from the fact that
neighboring \mnoVI\ octahedra share one oxygen atom) and weak second
and higher nn coupling.

We now address the issue of the unit-cell volume collapse observed
crystallographically~\cite{chatt;prb03} and seen in our Rietveld and
wide-range PDF refinements. Our Rietveld refinements show that the
average Mn-O bond length \rmnoave\ contracts across the transition
resulting a volume collapse of the \mnoVI\ octahedron, which is
found to fully account for the unit cell volume reduction. In the
absence of local structural information to the contrary, this
behavior is most readily explained if charges are delocalizing
resulting in more regular \mnoVI\ octahedra and a smaller unit cell,
as observed at higher doping in the La$_{1-x}$Ca$_x$MnO$_3$
system~\cite{billi;prl96}.  However, we have clearly shown that
\emph{locally} the \mnoVI\ octahedra do not change their shape and
the local $r_{Mn-O}$ is independent of temperature. The volume
collapse must therefore have another physical origin. The most
likely scenario is an increase in octahedral tilting amplitude. This
is similar to the behavior seen at the pseudo-cubic transition of
perovskite AlF$_3$~\cite{chupa;jacs04}.  If the increased tilting is
not long-range ordered a response is expected in oxygen displacement
parameters perpendicular to the Mn-O bond.  These are shown in
Fig.~\ref{fig;ouaniso}(b) as the circles.  The temperature
dependence indicates that this motion is rather soft ($U$ increases
strongly with temperature) but there is no clear discontinuity at
\tjt .
%%%%%%%%%%%%%%%%%%%%%%%%%%%%%%%%%%%%%%%%%%%%%%%%%%%%%%%%%%%%%%%%%%%%
%
%  Here FIGURE starts ...
%%%%%%%%%%%%%%%%%%%%%%%%%%%%%%%%%%%%%%%%%%%%%%%%%%%%%%%%%%%%%%%%%%%
%
%
%%%%%%%%%%%%%%%%%%%%%%%%%%%%%%%%%%%%%%%%%%%%%%%%%%%%%%%%%%%%%%%%%%%
%
%  Here FIGURE ends!
%%%%%%%%%%%%%%%%%%%%%%%%%%%%%%%%%%%%%%%%%%%%%%%%%%%%%%%%%%%%%%%%%%%
However, the change in tilt amplitude required to explain the volume
collapse is tiny. The average tilt angle needs to increase only by
0.4 degrees to account for the observed 0.38~\% volume collapse,
which is equivalent to displacing the oxygen atom by only $\sim$
0.014~\AA ,  thus we may not expect to see a discontinuity in $U$.

A large discontinuity in the oxygen displacement parameter
\emph{parallel} to the Mn-O bond is observed in the refinements
(Fig.~\ref{fig;ouaniso}(b) squares). This is precisely what would be
expected from a system with JT-distorted octahedra that are
orientationally disordered on average.

The sudden increase of conductivity across the JT
transition~\cite{manda;prb01,zhou;prb99} cannot be easily explained
by the locally JT distorted \mnoVI\ octahedra and increased
octahedral tilt in the high temperature phase. We speculate that
this conductivity anomaly comes along with the dynamic nature of the
JT distortions. Zhou and Goodenough have proposed a vibronic model
where some \mn{3+} ions charge disproportionate into \mn{2+} and
\mn{4+} \cite{zhou;prb99}. We see no direct evidence for this charge
disproportionation but if it occurs on a small minority of sites it
would undetectable in our data.

To summarize, the PDF clearly shows that the JT transition in
\lamno\ is of orbital order-disorder type; however, 16~\AA\ 
nanoclusters of short-range orbital order persist in the high
temperature $O$ and $R$-phases. Our analysis of the PDF shows that
this is a promising approach to extract nanocluster information in the
absence of single crystal diffuse scattering data.

%\acknowledgments
Work at MSU was supported by NSF through grant DMR-0304391. Work at
ANL was supported by DOE under Contract No.  W-31-109-ENG-38.
Beamtime on NPDF at Lujan Center was funded by DOE through contract
W-7405-ENG-36.

%---------------------------------------------------------------------------
% References
% NOTE: Add .bbl file here before submitting the file !!!!!!!!!!!!!!!!!!!!!!
%---------------------------------------------------------------------------

%
%
%    \bibliography{abb-billinge-group,%
%                  xiangyun,%
%                  simon,%
%                  billinge-group%
%    }
%
%    \bibliographystyle{aip_simon}
%
%---------------------------------------------------------------------------
% Figures and Tables (copy them here before submitting !!!!!!!!!!!!)
%---------------------------------------------------------------------------

\end{document}